# Possible Scale Invariant Linear Magnetoresistance in Pyrochlore Iridates $Bi_2Ir_2O_7$

**Jiun-Haw Chu[1,2,3,*], Jian Liu[1,2,4*], Han Zhang[4], Kyle Noordhoek[4], Scott C. Riggs[5,6], Maxwell Shapiro[5,6], Claudy Ryan Serro[7], Di Yi[5,7], M. Mellisa[1], S. J. Suresha[2], C. Frontera[8], E. Arenholz[9], Ashvin Vishwanath[1,2,+], Xavi Marti[7], I. R. Fisher[5,6], R. Ramesh[1,2,7]**

[1] Department of Physics, University of California, Berkeley, California 94720, USA
[2] Materials Science Division, Lawrence Berkeley National Laboratory, Berkeley, California 94720, USA
[3] Department of Physics, University of Washington, Seattle, Washington 98195, USA
[4] Department of Physics and Astronomy, University of Tennessee, Knoxville, Tennessee 37996, USA
[5] Department of Applied Physics and Geballe Laboratory for Advanced Materials, Stanford University, Stanford, California 94305, USA
[6] Stanford Institute of Energy and Materials Science, SLAC National Accelerator Laboratory, 2575 Sand Hill Road, Menlo Park 94025, California 94305, USA
[7] Department of Materials Science and Engineering, University of California, Berkeley, California 94720, USA
[8] Institut de Ciencia de Materials de Barcelona, ICMAB-CSIC, Campus UAB, E-08193 Ballaterra, Spain
[9] Advanced Light Source, Lawrence Berkeley National Laboratory, Berkeley, California 94720, USA
*E-mail: JHC: jhchu@uw.edu, JL: jianliu@utk.edu

+Current address: Department of Physics, Harvard University. Cambridge, Massachusetts

### Abstract

We report the observation of a linear magnetoresistance in single crystals and epitaxial thin films of the pyrochlore iridate $Bi_2Ir_2O_7$. The linear magnetoresistance is positive and isotropic at low temperatures, without any sign of saturation up to 35 Tesla. As temperature increases, the linear field dependence gradually evolves to a quadratic field dependence. The temperature and field dependence of magnetoresistance of $Bi_2Ir_2O_7$ bears strikingly resemblance to the scale invariant magnetoresistance observed in the strange metal phase in high $T_c$ cuprates. However, the residual resistivity of $Bi_2Ir_2O_7$ is more than two orders of magnitude higher than the curpates. Our results suggest that the correlation between linear magnetoresistance and quantum fluctuations may exist beyond high temperature superconductors.

Keywords: Pyrochlore iridates, linear magnetoresistance

## 1. Introduction

The interplay between conduction electrons and symmetry breaking phases has been a central problem in condensed matter physics. One recent example is the rare earth pyrochlore iridate ($A_2Ir_2O_7$, A = rare earth or Yttrium). It has attracted great attention in the past few years because of its potential to host exotic phases that intertwine non-trivial band topology with magnetic order [1,2]. Several theoretical proposals suggest the realization of strongly correlated topological phases in both the bulk and thin films[1-5]. Experimentally, it was found that the ground state can be tuned by the A site rare earth ions, from the magnetically ordered insulator $Y_2Ir_2O_7$ to the magnetically disordered metal $Pr_2Ir_2O_7$ [6]. In particular, exotic magneto-transport behavior was observed in several magnetically ordered pyrochlore iridates with the all-in-all-out configuration [7-10]. Nevertheless, most of the studies concentrate on the magnetically ordered state and less attention was given to the magnetic fluctuations in the disordered phase.

Here we focus on a sister compound $Bi_2Ir_2O_7$, in which $R^{3+}$ of rare earth pyrochlore iridate is replaced by $Bi^{3+}$ ions. From the early studies $Bi_2Ir_2O_7$ appears to be a featureless metal, with an almost temperature independent resistivity[11]. Nevertheless, its magnetic susceptibility exhibits a Curie-Weiss temperature dependence with a Weiss temperature very close to zero, a behavior that is not expected from a conventional metal. This Curie-Weiss susceptibility was initially attributed to magnetic impurities, but it was repeatedly observed in several later studies[12,13], including measurements performed on single crystals [14]. In addition, μSR study revealed a possible magnetic transition occurs at a temperature below 1.8K [15]. All these experimental findings suggest that $Bi_2Ir_2O_7$ might be close to a magnetic quantum phase transition in the global phase diagram of pyrochlore iridates, and the associated fluctuations could strongly influence its electronic properties.

In this work we study the interplay between magnetic fluctuations and conducting electrons by magnetoresistance (MR) measurement of epitaxial thin films and single crystals of $Bi_2Ir_2O_7$. We observe a positive isotropic linear MR at low temperatures, with no sign of saturation up to 35 Tesla. The MR shows a strong temperature dependence, unlike the temperature independent zero-field resistivity. Intriguingly, the field and temperature dependence of MR and the linear slope of field dependence show striking similarity between $Bi_2Ir_2O_7$ and optimally doped high Tc cuprates. We discuss our results in the context of possible existence of scale invariant MR in a highly disordered metal.

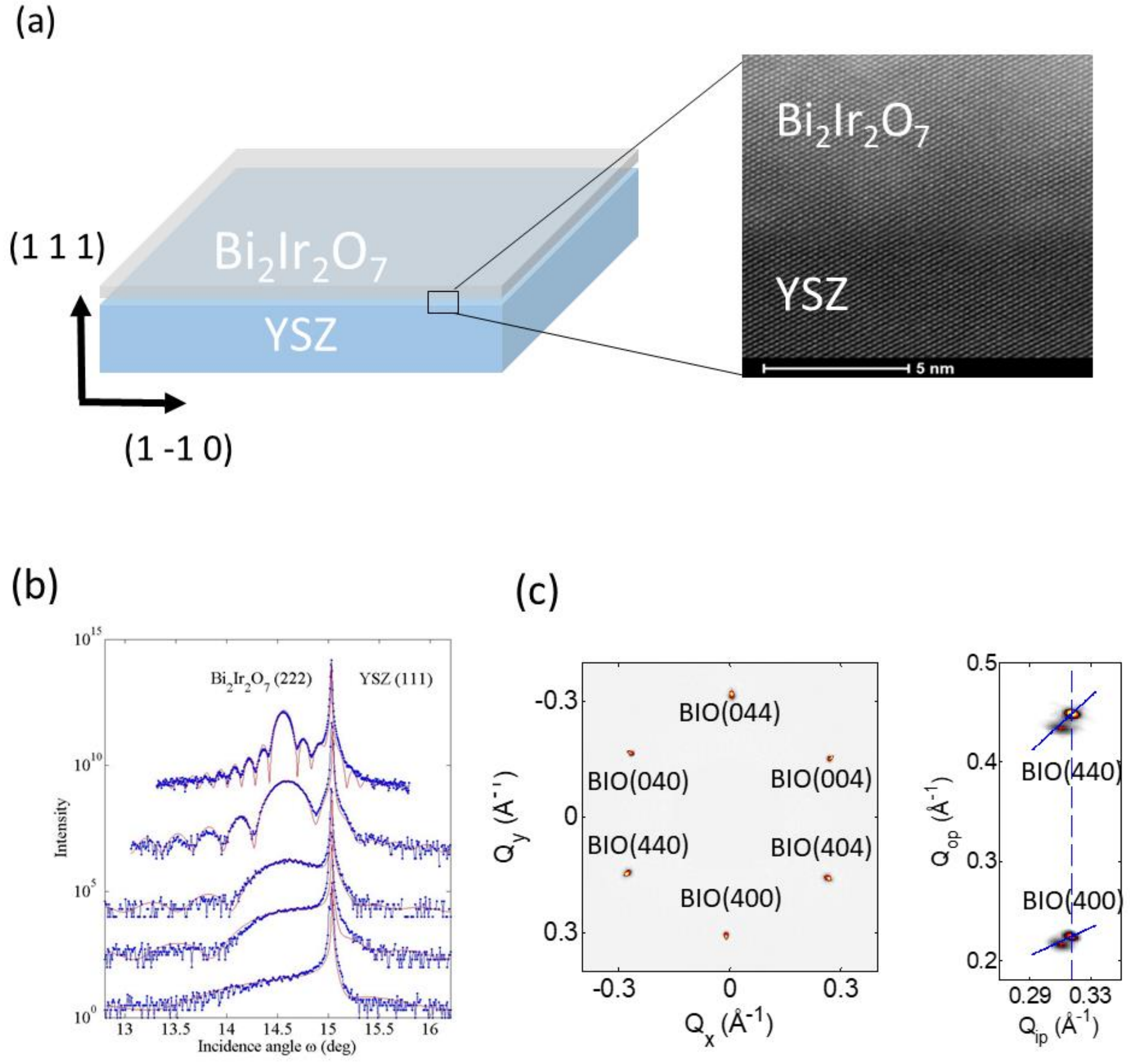

**Fig. 1 Basic structure characterization of epitaxial BIO/YSZ(111) thin films**. (a) Left: Schematics of substrate and thin film along with the crystallographic orientations. Right: Cross-section TEM image. (b) Specular ω-2θ scan of x-ray diffraction for 3 to 30nm thick films, showing the (111) and (222) peaks of the substrate and tin film respectively. The red solid curves show fitting described in the main text. (c) (Left) Pole figure around the (400) and (440) reflections of $Bi_2Ir_2O_7$ (BIO) displaying the three-fold in plane symmetry of the (111)-textured films. (Right) Reciprocal space maps around the BIO(400) and BIO(440) reflections; the strong peaks correspond to YSZ substrate reflections (200) and (220), respectively. Dashed line indicates the location of the substrate. Solid lines signal the fully-relaxation line of a cubic structure. Data shows that BIO films are not relaxed but are under tension in-plane.

## 2. Methods

Single crystals of $Bi_2Ir_2O_7$ are grown from a binary melt of $B_2O_3$ and $IrO_2$, the details of growth and characterization are described elsewhere[16]. Epitaxial thin films of $Bi_2Ir_2O_7$ were grown on (111)-oriented yttria-stabilized zirconia (YSZ) substrates using pulsed laser deposition. A KrF excimer laser (248 nm wavelength) was used at a repetition rate of 1 Hz. The laser beam was focused to fluency of 0.75 $Jcm^2$ on a stoichiometric $Bi_2Ir_2O_7$ target; the substrate placed at a distance of 7.5 cm. The films were deposited at a substrate temperature of 550 °C and a 50 mTorr of oxygen background pressure. At the end of the growth, the samples were cooled down in 1 atm of oxygen pressure. X-ray diffraction patterns revealed a good agreement between experiments and theoretical modeling, from which we obtained the thickness and lattice parameters of the layers. The fitting rendered an interface root-mean-square roughness below one nanometer, and a good control on film's thickness down to 3nm [17]. Exhaustive reciprocal space mapping in Fig. 1(c) revealed an in-plane three-fold symmetry with no traces of spurious phases or orientations other than the principal (111)-texture of the films.

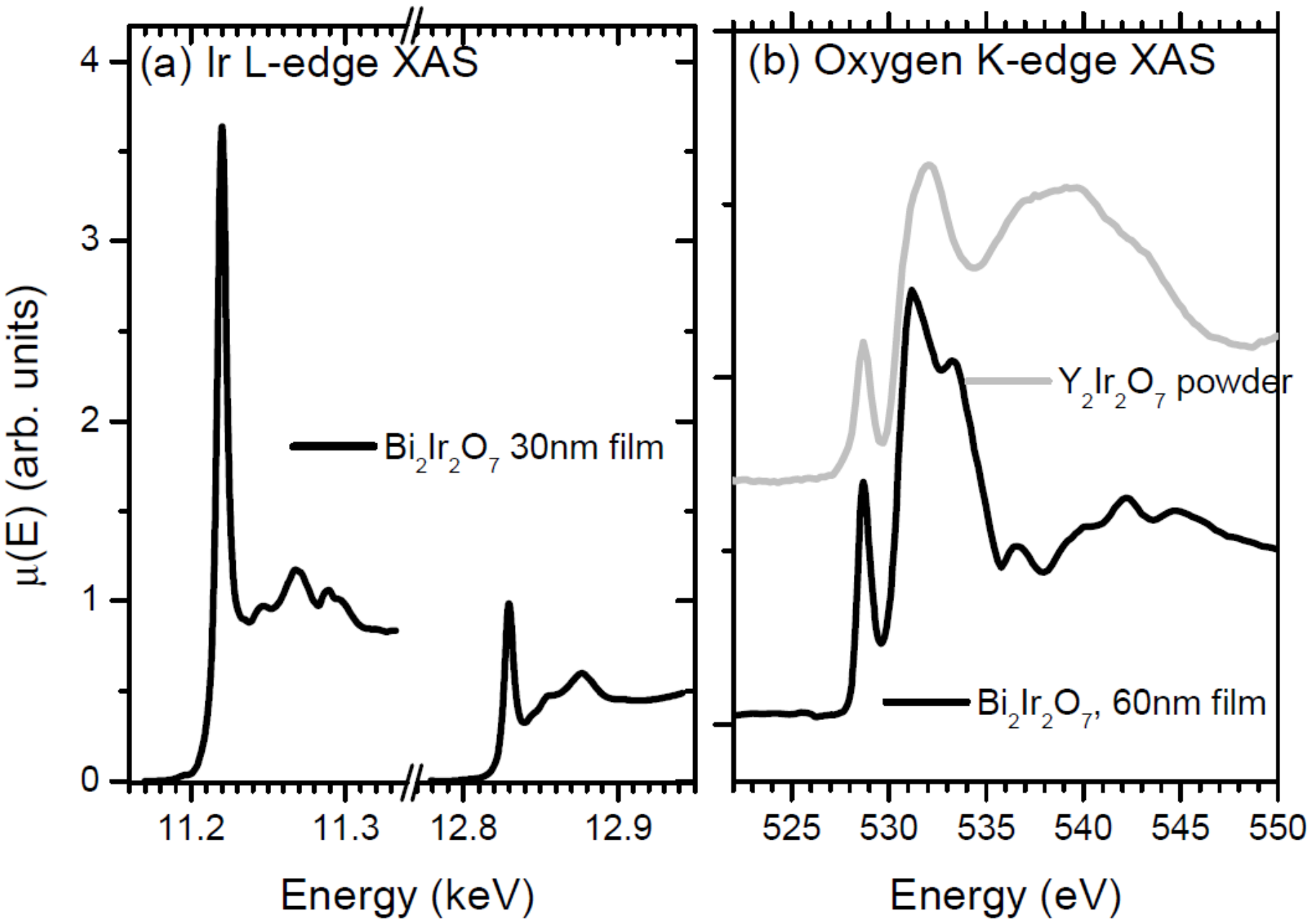

**Fig. 2 X-ray absorption characterization of $Bi_2Ir_2O_7$ films.** (a) X-ray absorption spectra collected at the Ir L3 (E = 11.215 keV) and L2 (E = 12.824 keV) absorption edges for a 30nm thick $Bi_2Ir_2O_7$ film. (b) X-ray absorption spectra collected at the Oxygen K absorption edge for a 60nm thick $Bi_2Ir_2O_7$ film. The spectrum of the powder sample of $Y_2Ir_2O_7$ is also plotted for comparison. The curves are shifted vertically for clarity.

X-ray absorption spectrum collected at Ir $L_{3,2}$-edge for a 30nm $Bi_2Ir_2O_7$ film is presented in Fig. 2(a). There are two important quantities that could be extracted from this spectrum: the position of "white line" feature providing the valence state of iridium and the branching ratio $BR = I_{L2}/I_{L3}$ ($I_{L3,2}$ is the integrated absorption intensity) that quantifies the degree of spin-orbit coupling. We compared the white line peak position with that obtained during the same experiment from the reference $Sr_2IrO_4$ [18], the peak positions are identical within experimental resolution (~ 0.2eV), which confirms that Ir in $Bi_2Ir_2O_7$ is in the 4+ state. The branching ratio $BR$ is related to $< \mathrm{l} \times \mathrm{s} >$ and the number of d-holes $\mathrm{n_h}$ as $\mathrm{BR} = (2 + \mathrm{r})/(1 - \mathrm{r})$ with $\mathrm{r} = < l \cdot s >/< \mathrm{n_h} >$. This is due to the dipole selection rules of the absorption transition from the spin-orbitally split $2p_{3/2}$ and $2p_{5/2}$ core levels [19]. We followed the same analysis procedure used in ref [20] and obtained a BR =5.6 for $Bi_2Ir_2O_7$, which is equivalent to $< l \cdot s > = 2.7$. These numbers are comparable to those obtained from the other iridates compound, suggesting that Ir 5d states in $Bi_2Ir_2O_7$ can indeed be described by the $d^5$ low-spin state with a half-filled $J_{eff} = 1/2$ state at the Fermi level. The measured $< l \cdot s > =$ 2.7 can then be decomposed into contributions from the single $J_{eff} = ½$ hole ($< l \cdot s > \approx 1$) and the four $e_g$ holes ($4 \times \frac{3}{2}\frac{\lambda}{10Dq} \approx 1.7$), where $\lambda$ and $10Dq$ are the spin-orbit coupling coefficient and the octahedral crystal field splitting, respectively.

X-ray absorption spectra of $Bi_2Ir_2O_7$ thin films and $Y_2Ir_2O_7$ powders collected at oxygen K-edge is also presented in Fig. 2(b). Since oxygen hybridizes with every element of the compounds the oxygen K-edge excitation is generally regarded as a measure of joint density of states between these elements and oxygen above Fermi level. The absorption spectrum right above the edge is characterized by a narrow peak from the $J_{eff}$ = 1/2 state at 528.5 eV. The pre-peak intensity of $Bi_2Ir_2O_7$ is relatively stronger than $Y_2Ir_2O_7$, probably due to stronger hybridization, consistent with the metallic ground state of $Bi_2Ir_2O_7$. At higher energies, and a broad structure from 529 to 534 eV due to the broad $e_g$ band. The spectral shape of this feature is modified to be a double-peak structure in $Bi_2Ir_2O_7$, which may be caused by the relatively low energy of the 6sp states of Bi overlapping with the top of the valence band. Band structure calculations using local density approximation (LDA) on density functional theory (DFT) suggested that in addition to the $J_{eff}$= 1/2, Bi 6p states also contributed to the sates close to Fermi level and significantly broadening the bandwidth[14]. The similarity of the pre-peak region with no clear broadening between $Bi_2Ir_2O_7$ and other rare earth pyrochlore iridates indicates that the electronic structure has not been significantly altered by the Bi 6p sates.

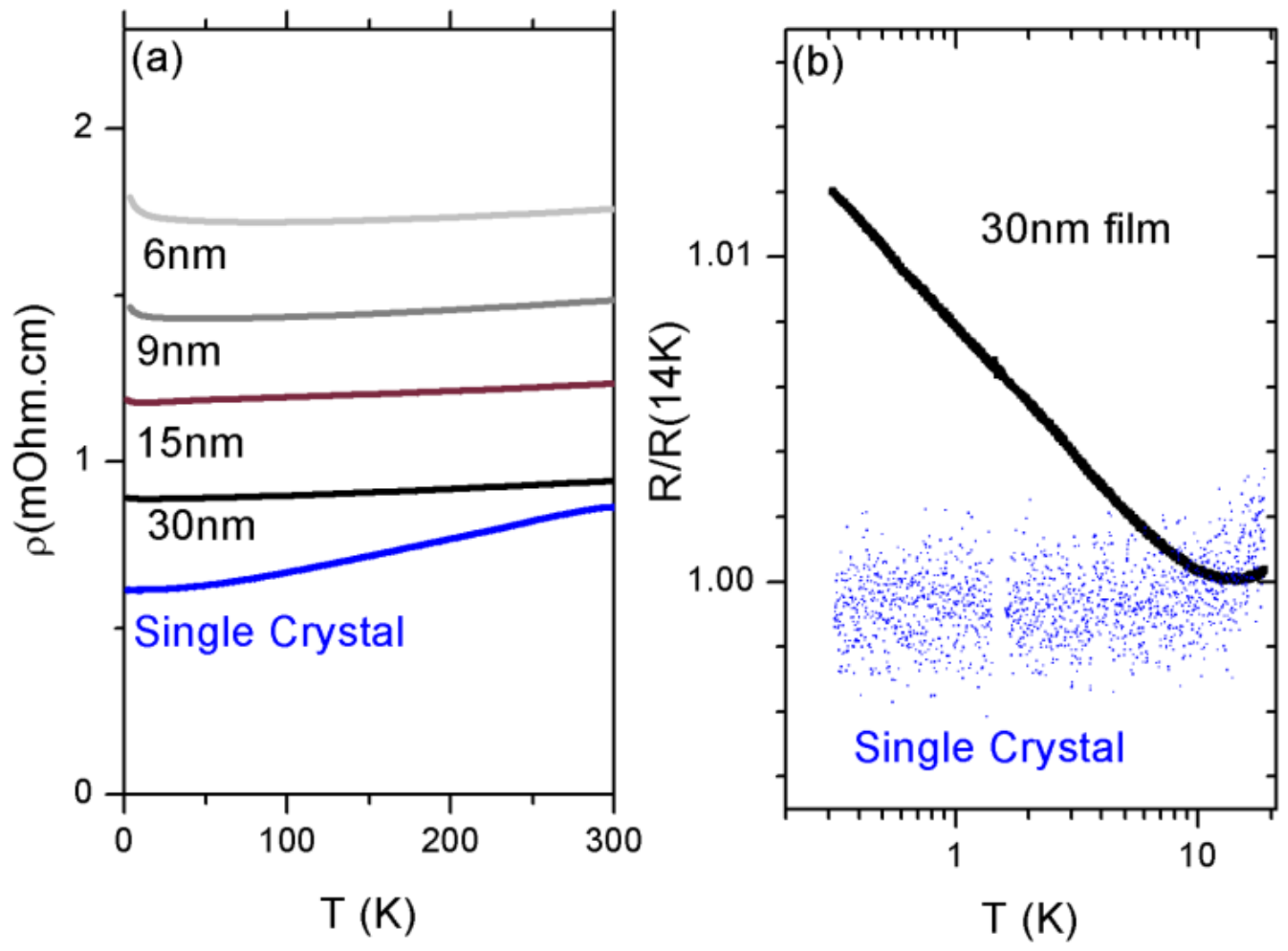

**Fig. 3:** Temperature dependence of resistivity for thin films and single crystals of $Bi_2Ir_2O_7$. (a) Resistivity versus temperature of a single crystal (blue) and 30nm to 6nm thick films (black to grey). (b) Semi-log plot of resistance vs temperature of a 30nm thin film (black) and a single crystal (blue) for the temperature range T = 0.32 to 20K. Resistance is normalized by the value at T = 14K.

Electrical transport measurements were performed at the National High Magnetic Field Laboratory (NHMFL) in Tallahassee in DC magnetic fields up to 35 T and in a Physical Properties Measurement System (PPMS) in fields up to 14 T. Unless otherwise noted, the magnetic field was always oriented perpendicular to the film plane, and parallel to the [111] direction for single crystals. Measurements were made for both positive and negative field orientations and only the even component is extracted for MR.

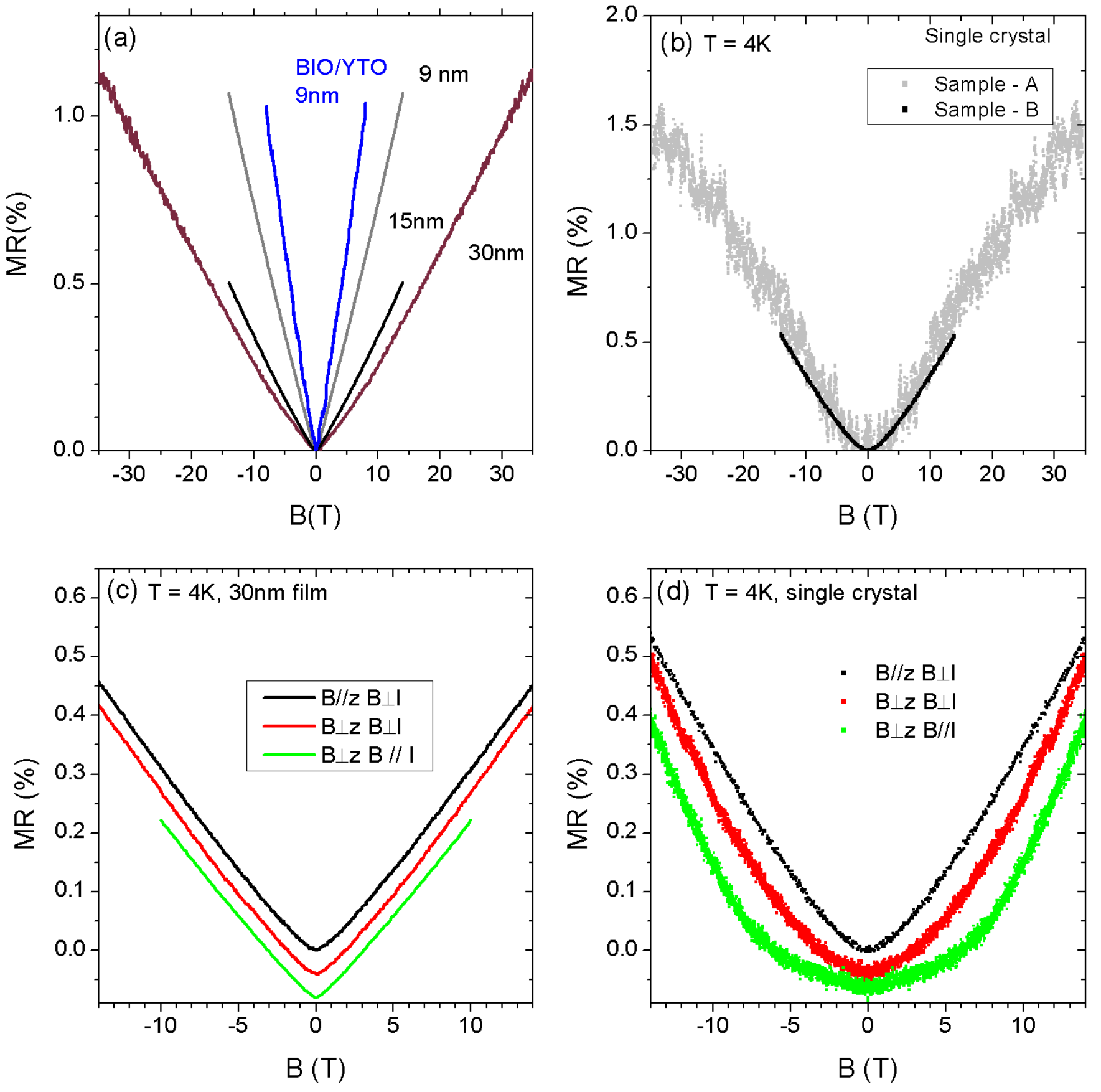

**Fig. 4: Angle dependence of magnetoresistance at low temperatures.** (a) Magnetoresistance of 9nm to 30nm thick films at T = 4K. The YTO-buffered 9nm BIO film was measured at T = 2K. Magnetic fields were applied along the out-of-plane (111) direction. (b) Magnetoresistance of two single crystal samples. Magnetic fields were applied parallel to (111) direction and perpendicular to the current direction. (c)(d) Magnetoresistance of (c) a 30nm thick film and (d) a single crystal at T = 4K for three field-current configurations: fields out-of-plane(black), in-plane perpendicular to the current(red) and in-plane parallel to the current(green). Data are offset by 0.05% for clarity.

## 3. Results

The resistivity versus temperature for both single crystal and thin films of $Bi_2Ir_2O_7$ is shown in Fig. 3(a). They both host an almost temperature independent resistivity, with a residual resistance ratio (RRR) 1.4 for the single crystal and 1.1 for the thin film samples, consistent with previous measurements. The room temperature resistivity is about 1 mΩcm. For thin films the resistivity increases as the film thickness decreases, which is likely due to the increase of surface scattering. Using the Fermi wave vector $k_F = 0.3\text{Å}^{-1}$ extracted from the photoemission experiment[21], the mean free path of $Bi_2Ir_2O_7$ is about 1-2nm, just above the Mott-Ioffe-Regel limit.

As the temperature decreases the resistivity of thin films reaches a minimum and shows an upturn. The resistivity minimum appears in all films, and the temperature of which increases as the film thickness decreases. Fig. 3(b) shows the low temperature resistivity of a 30nm film and a single crystal in a log temperature scale, and it can be clearly seen that the upturn in thin film sample follows a logarithmic temperature dependence, a signature of weak localization or disorder enhanced electron-electron interaction in two-dimensions. This is expected from the short mean free path and low RRR; the transport of $Bi_2Ir_2O_7$ is highly diffusive and the electrons are mainly scattered by static disorder.

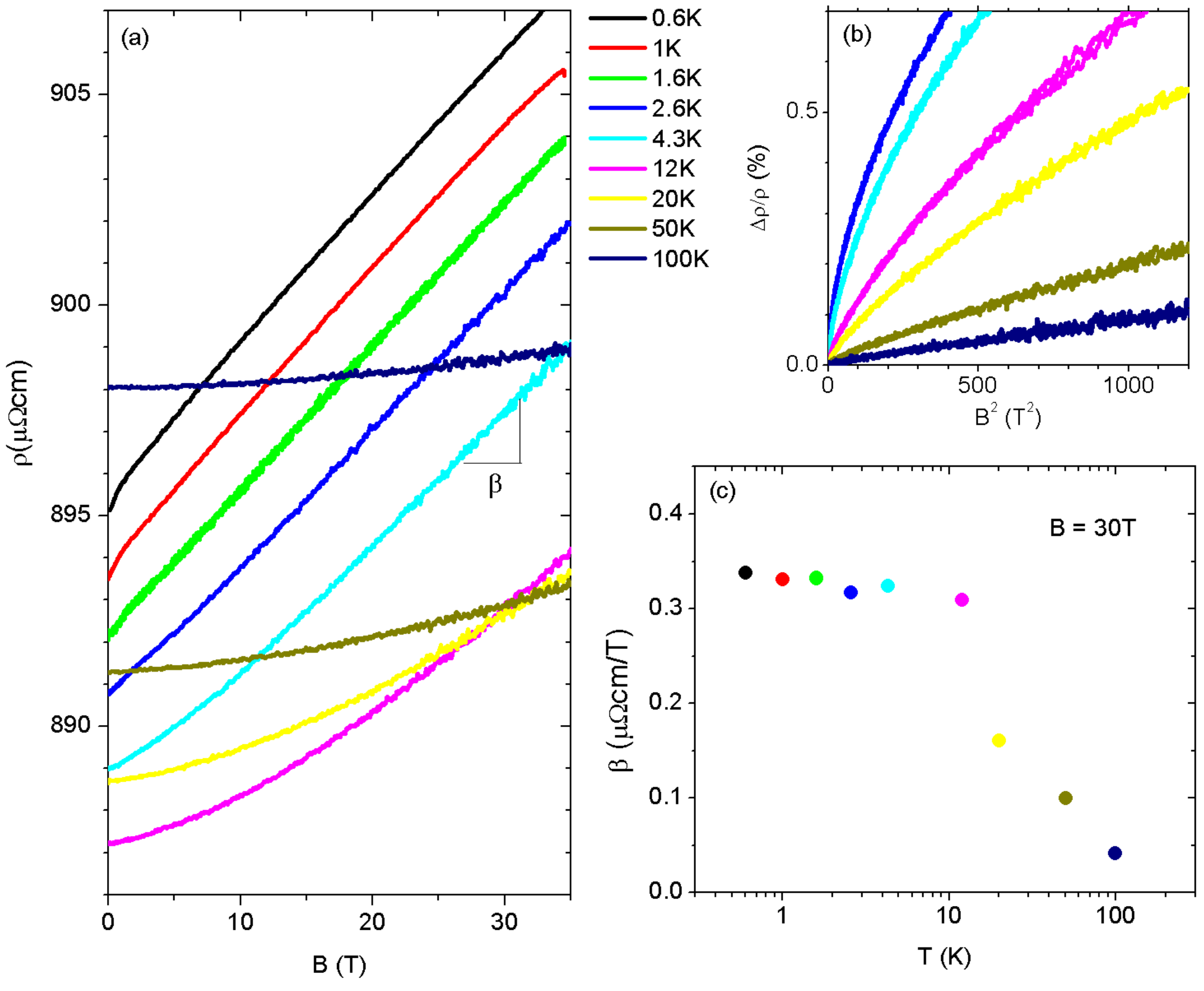

**Fig. 5** Temperature dependence of magnetoresistance: (a) Resistivity as a function of field of a 30nm thick film measured at $T = 0.6\text{ to }100\text{K}$. Magnetic fields were applied perpendicular to the film plane. (b) Magnetoresistance for T > 2.6K plotted in $B^2$. (c) The temperature dependence of high field linear slope of magnetoresistance $\beta = d\rho/dB$ at B = 30T.

In contrast to the featureless zero-field resistivity, the MR of $Bi_2Ir_2O_7$ exhibits an unusual field and temperature dependence. The MR of single crystals and films at T = 4K is summarized in Fig. 4. The MR of films of different thickness is plotted in Fig. 4(a), which shows a clear linear field dependence. This is in agreement with the previous report [22]. To verify whether this is an intrinsic effect, we also grew a 9nm film with a 20nm $Y_2Ti_2O_7$ (YTO) buffer layer, which shares the same pyrochlore structure as BIO and eliminates the BIO/YSZ interfacial region. One can see this reference sample indeed shows a qualitatively similar behavior. Fig. 4(b) plots the MR of single crystal samples, which also shows a similar linear field dependence at high field. Notice that for the 30nm film and for the single crystal (sample A) the MR was measured up to 35T, with no sign of saturation of the linear field dependence. We note that the signal to noise ratio is much better for the thin film sample than the single crystal. This is due to the much higher resistance of the thin film sample. Despite the resistivity is similar for single crystals and thin films, the resistance of the single crystal sample is typically 1mΩ, whereas the resistance of thin films is typically greater than 100Ω. This difference is coming from the sample geometric shape factor.

We also measured the angle dependence of MR. The field was applied along three directions: out of the plane, within the plane but perpendicular to the current, and within the plane and parallel to the current, as shown in Fig. 4(c) and (d). The MR measured in different configurations is essentially identical for thin films (Fig. 4(c)), the curves perfectly overlap with each other if there is no offset. The MR of the single crystal is also highly isotropic (Fig. 4(d)), with only a small difference in lower field range. In general, the orbital contribution to MR is greatest when the field is perpendicular to the current and vanishes when they are parallel. The highly isotropic nature of linear MR strongly suggests that the origin of it is related to the spin degree of freedom rather than the orbital motion.

Having established that the MR of BIO is linear and isotropic at helium temperatures, we present its temperature dependence. We focus on the data taken from 30nm thin film sample because of its superior signal to noise ratio and the MR of single crystals show qualitatively similar behavior. Fig. 5 (a) summarizes the resistivity as a function of field measured between T = 0.6 to 100K with field up to 35T. The field dependence of resistivity is essentially the same between 0.6 and 2.6K with a linear field dependence and a temperature independent slope. As temperature increases above 4K the low field MR begins to deviate from linear, but the high field MR remains linear. At even higher temperatures, the field dependence of MR gradually evolves from linear to quadratic, which can be seen more clearly in Fig. 5 (b) in which the data are plotted against $B^2$. To gain more insight, we plot the slop of MR $\beta = d\rho/dB$ at B = 30T as a function of temperature. As shown in Fig. 5 (c), $\beta$ saturates to a value of 0.33 $\mu\Omega\mathrm{cm/T}$ below 10K. The temperature dependence of $\beta$ suggests that the temperature sets an energy scale. When magnetic field surpasses this energy scale the MR recovers the linear field dependence.

## 4. Discussion

The combination of highly isotropic nature, strong temperature dependence and low RRR makes this linear MR very difficult to be explained by several conventional theories. It’s certainly not coming from the semi-classical Lorentz mechanism, which is generally sensitive to the angle between field and current directions. In addition, the violation of the Kohler's rule is evident in that the MR increases by more than tenfold as temperature decreases while the zero-field resistance is almost the same. The explanation based on Abrikosov’s quantum linear MR is also unlikely [23], since it requires small Fermi pockets with high mobility carriers where as $Bi_2Ir_2O_7$ has large Fermi surfaces with a short mean free path. The explanation based on classical linear MR in disordered system is also not applicable for the same reason, since it relies on mobility fluctuations that only exist in low carrier density semiconductors and semimetals [24,25].

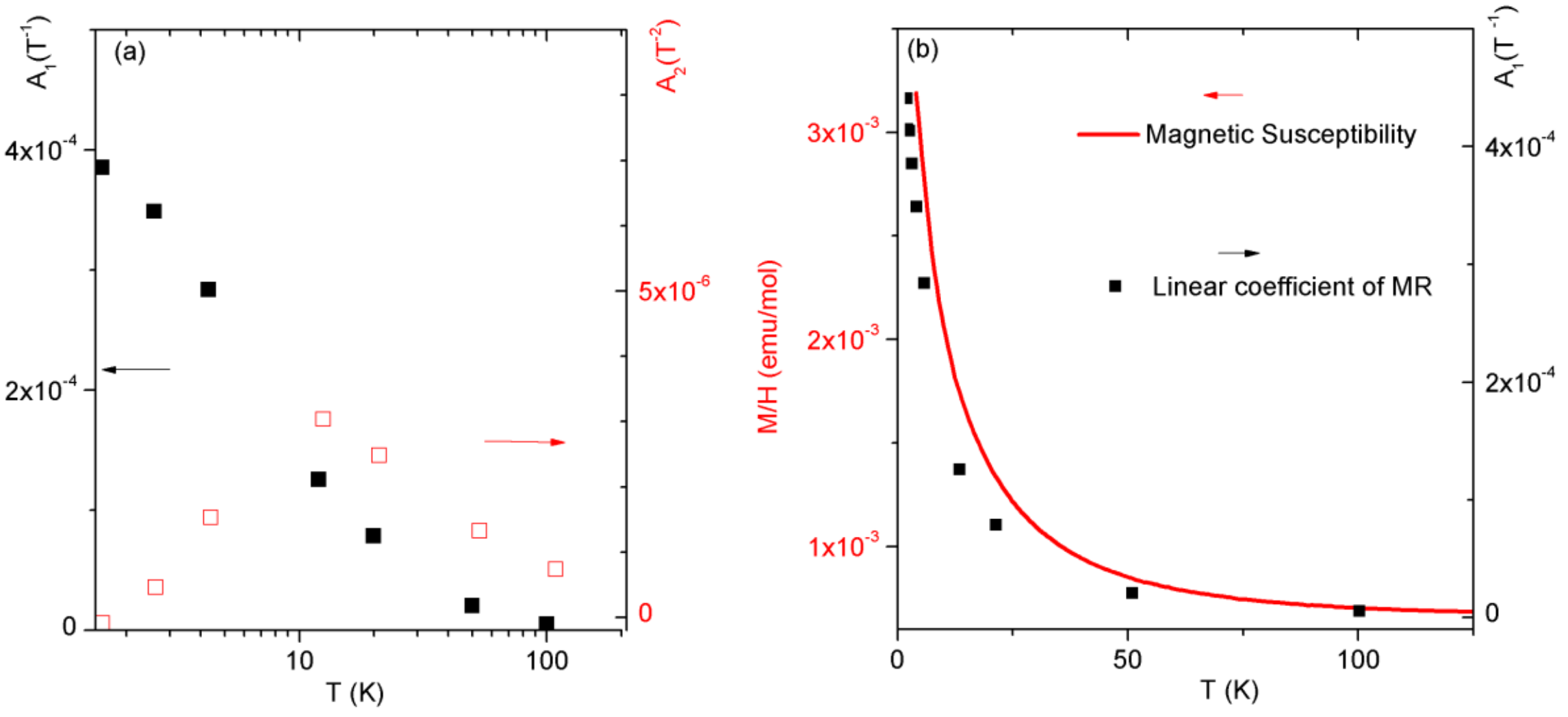

**Fig. 6** (a) The extracted linear ($A_1$, left) and quadratic ($A_2$, right) coefficients of magnetoresistance as a function of temperature. (b) Magnetic susceptibility (*M/H*) of the single crystals $Bi_2Ir_2O_7$, measured at 5000 Oe (left axis), and the linear coefficient $A_1$ (right axis) as a function of temperature.

The fact that the MR is almost identical for all field-current configurations strongly suggests that it is a spin related phenomenon. Although a positive, linear isotropic MR does exist in highly disordered ferromagnets, a spontaneous magnetization is a necessary ingredient[26,27]. There is no sign of long-range magnetic order above $T = 1.8K$ in $Bi_2Ir_2O_7$, yet the magnetic susceptibility shows a Curie-Weiss temperature dependence. In the absence of local moments in the metallic system, the Curie-Weiss susceptibility indicates the existence of strong magnetic fluctuations that persist to high temperatures.

To gain more insight, we fit the normalized MR to a second order polynomial ($\Delta R/R = A_1|B| + A_2B^2$) [28]. The temperature dependence of the extracted linear and quadratic coefficients $A_1$ and $A_2$ are plotted in Fig. 6 (a). At $T = 1.6K$, the linearity of MR is confirmed by the vanishingly small quadratic coefficient $A_2$ compared to linear component $A_1$. As temperature increases, the quadratic coefficient $A_2$ increases, whereas the linear coefficient $A_1$ decreases. Plotting the linear coefficient $A_1$ of MR and the magnetic susceptibility together in Fig. 6 (b), we found striking similarity in the temperature dependence. This empirical observation strongly suggests that these two quantities may share a common origin., i.e. diverging quantum magnetic fluctuations.

Recently, there are several reports of the observation of scale invariant linear magnetoresistance close to the quantum critical points in high temperature superconducting cuprates and iron pnictides [29-32]. From a phenomenological point of view, the MR of $Bi_2Ir_2O_7$ shows striking similarity to these two systems: the resistivity depends linearly on field as temperature approaches zero, and the zero-field temperature dependence of resistivity also becomes linear at high temperatures. If we compare the linear slope of MR and the linear slope of temperature dependence of resistivity in natural units, the two quantities $\beta/\mu_B$ = 5.7 $\mu\Omega cm/meV$ and $\alpha/k_B$ = 2.9 $\mu\Omega cm/meV$ are comparable, which is also the case for cuprates and pnictides. In fact, the value of $\beta/\mu_B$ = 5.7 $\mu\Omega cm/meV$ in $Bi_2Ir_2O_7$ is very close to what was observed in cuprates (5.2 $\mu\Omega cm/meV$)[30]. All these comparisons suggest that the linear MR in $Bi_2Ir_2O_7$ may be a consequence of its proximity to quantum critical points.

Despite the striking similarity, we close our discussion by pointing out a major difference between $Bi_2Ir_2O_7$ and high Tc cuprates, i.e. the degree of disorder. The residual resistivity is about 2 to 3 orders of magnitude higher in $Bi_2Ir_2O_7$, which means that all the temperature and field dependence of resistivity constitutes only about one percent of total resistivity. It's unclear in the situation the temperature and field dependence of resistivity and residual resistivity can be cleanly separated, since the Matthiessen's rule may breakdown in the strong disorder limit. It's also unclear how the disordered enhanced quantum interference may affect the low temperature MR. Recently, the scale invariant linear MR and quantum critical behavior have also been reported in highly disordered metals[33,34]. A systematic study as a function of disorder will offer valuable insight into this challenging and interesting problem.

## 5. **Conclusion**

In summary, despite the large residual resistivity, the behavior of MR $Bi_2Ir_2O_7$ shows striking similarity to the strange metal phase in cuprates both qualitatively and quantitatively, which is consistent with it's being close to a magnetic quantum phase transition. Our work present $Bi_2Ir_2O_7$ as another platform to study the intriguing phenomena of quantum critical metal in the strong disorder limit.

## Acknowledgement

We thank the experimental support by D. Graf. This work is supported by the director of the Office of Science, Office of Basic Energy Sciences, Materials Sciences and Engineering Division, of the U.S. Department of Energy under Contract No. DE-AC02-05CH11231 through the Quantum Materials FWP. Use of the Advanced Photon Source, an Office of Science User Facility operated for the U.S. Department of Energy (DOE) Office of Science by Argonne National Laboratory, was supported by the U.S. DOE under Contract No. DE-AC02-06CH11357. A portion of this work was performed at the National High Magnetic Field Laboratory, which is supported by the National Science Foundation Cooperative Agreement No. DMR-1157490, the State of Florida and the United States Department of Energy. $Bi_2Ir_2O_7$ crystals and targets were synthesized and characterized at Stanford University supported by the Department of Energy, Office of Basic Energy Sciences under contract DE-AC02-76SF00515. J.L. acknowledges support from the Organized Research Unit Program at the University of Tennessee. J.-H. C. also acknowledge the support from Alfred P. Sloan Foundation.